%Paper: hep-th/9305034
%From: sasakura@danjuro.phys.s.u-tokyo.ac.jp (naoki sasakura)
%Date: Mon, 10 May 93 17:59:35 +0900

%
%Plain Tex file , input phyzzx
%

\input phyzzx

\pubnum{UT-642}

%%%%%%%%%%%%%%%%%%%%%%%%%%%%%%%%%%%%%%%%%%%%%%%%%%%%%%%%%%%%%%%%%%%%

\def\PRL{Phys.~Rev.~Lett.}
\def\CMP{Commun.~Math.~Phys.}
\def\PL{Phys.~Lett.}
\def\MPL{Mod.~Phys.~Lett}
\def\NP{Nucl.~Phys.}

%%%%%%%%%%%%%%%%%%%%%%%%%%%%%%%%%%%%%%%%%%%%%%%%%%%%%%%%%%%%%%%%%%%%

\REF\JW{J.F.~Wheater \journal \PL &B223 (89) 451.}

\REF\DW{R.~Dijkgraaf and E.~Witten \journal \CMP &129 (90) 393.}

\REF\KM{Al.R.~Kavalov and R.L.~Mkatchyan \journal \PL &242B (90)
429.}

\REF\TJ{T.~J{\`o}nsson \journal \PL &B265 (91) 141.}

\REF\TV{V.G.~Turaev and O.Y.~Viro, ``{\it State Sum Invariant of
3-Manifolds
and Quantum 6j-Symbols},'' preprint (1990).}

\REF\OS{H.~Ooguri and N.~Sasakura \journal \MPL &A6 (91) 3591.}

\REF\MT{S.~Mizoguchi and T.~Tada \journal \PRL &68 (92) 1795.}

\REF\AW{F.~Archer and R.M.~Williams \journal \PL &B273 (91) 438.}

\REF\DJN{B.~Durhuus, H.~Jacobsen and R.~Nest \journal \NP
(Proc.~Suppl.) &25A (92) 109.}

\REF\BOU{D.V.~Boulatov \journal \MPL &A7 (92) 1629.}

\REF\HO{H.~Ooguri \journal \MPL &A7 (92) 2799.}

\REF\FJA{F.J.~Archer \journal \PL &B295 (92) 199.}

\REF\BP{C.~Bachas and P.M.S.~Petropoulos \journal \CMP &152 (93)
191.}

\REF\FHK{M.~Fukuma, S.~Hosono and H.~Kawai, preprint CLNS~92/1173.}

\REF\NN{B.~Nienhuis and M.~Nauenberg \journal \PRL &35 (75) 477.}

%%%%%%%%%%%%%%%%%%%%%%%%%%%%%%%%%%%%%%%%%%%%%%%%%%%%%%%%%%%%%%%%%%%

\titlepage

\title{Lattice Topological Field Theory and First Order Phase
Transition
\footnote{*}{Work supported by the Grant-in-Aid for Scientific
Research
from the Ministry of Education No.~04-1324.}}

\author{Naoki Sasakura
\footnote{\star}
{E-mail address: SASAKURA@TKYUX.PHYS.S.U-TOKYO.AC.JP}
\footnote{\dagger} {A Fellow of the Japanese Society for the
Promotion of Science
for Japanese Junior Scientists.}
\address{Department of Physics, University of Tokyo, Tokyo, 113,
Japan}}

\abstract{ Carrying out perturbations around a lattice topological
field theory in two dimensions, we show
that it is
on a first order phase transition fixed point with multiplicity
${n(n-1)/2}$, where $n$ is the number of its independent physical
observables.
We discuss about the order parameters and the finite size effect for
the free energy.
The finite size effect is described by the topological field theory.
We investigate also the renormalization group flow near the fixed
point,
and show that the flow agrees with that of the
Nienhuis-Nauenberg criterion.}

\endpage

It is a well known fact that a continuum field thoery can be
defined on a lattice along a renormalization group flow
where the limit is a fixed point on
a second order phase transition surface.
The correlation length diverges on the fixed point.
On the other hand there is another possibility for a fixed point of
a renormalization group flow, where the correlation length vanishes.
Are there any field theories corresponding to
such a zero-correlation-length fixed point?
Since the correlation length vanishes on such a fixed point,
we expect that the correlation functions of some local
observables are independent of their positions.
This would be a characteristic nature
of topological field theories.
Recently some lattice topological field theories have been
constructed\refmark{\JW-\FHK}.
The specific nature of these lattice theories is that their physical
observables are
completely independent of the lattice structures on which they are
defined.
In this letter we investigate how the system is influenced by
small changes of the local weights of the lattice
topological field theories in two
dimensions\refmark{\JW,\TJ,\BP,\FHK},
and investigate what kind of points they reside on.

The basic elements of the two-dimensional topological lattice
theory are
the tensor $C^T_{ijk}$ and the metric $g^{ij}$ of complex numbers
satisfying the following constraints\refmark{\FHK}:
$$
\eqalign{
C^T_{ijp}g^{pp'}C^T_{p'kl}=C^T_{jkp}g^{pp'}C^T_{p'li}, \cr
g^{ij}=g^{ii'}g^{jj'}C^T_{i'kl}g^{kk'}g^{ll'}C^T_{j'l'k'},\cr
}
\eqn\con
$$
where the Roman indices run from 1 to $M$, and the tensor and the
metric are invariant
under the cyclic permutations of the indices.
Consider a triangulated two-dimensional oriented surface with
genus $g$.
We assign
$C^T_{ijk}$ to each triangle, where $i,j,k$ represent the degrees of
freedom on its
three edges. And we assign $g^{ij}$ to each edge
and connect the tensors on the two triangles which have the
common edge. Then we will obtain a numerical value for the
triangulated surface
and call it the partition function \refmark{\JW,\TJ,\BP,\FHK} of
the surface:
$$
Z^T_g\equiv C^T_{i_1 i_2 i_3} g^{i_3 i_4} \cdots.
\eqn\toppart
$$
The condition \con\ guarantees the invariance of the partition
function
\toppart\ under the Mateev moves of the triangulation\refmark{\FHK},
and hence it is
independent of the triangulation of the
surface\refmark{\JW,\TJ,\BP,\FHK}.

We make a small shift of the tensor $\delta C_{ijk}$
from the original value corresponding to the topological theory:
$$
C_{ijk}=C^T_{ijk}+\delta C_{ijk}.
$$
We define the partition function by substituting $C^T_{ijk}$
with $C_{ijk}$ in \toppart.
Since the tensors do not necessarily satisfy the condition \con\
anymore,
the partition function depends on the triangulation.
Thus from now on we consider certain triangulations of the surface.
We will calculate the partition function
perturbatively in $\delta C_{ijk}$.
The first order terms are given by
$$
Z_g^{(1)}=\sum_{x} \delta C_{ijk} \langle P^{ijk}(x) \rangle_g,
\eqn\prt
$$
where the operator $P^{ijk}(x)$ denotes the puncture
operator\refmark{\FHK} of the triangle
at $x$ with indices $i,j,k$ on its edges, and the one-point function
$\langle P^{ijk}(x) \rangle_g$ is that of
the original lattice topological field theory on the surface with
genus $g$.
Because of the topological nature, the one point function
$\langle P^{ijk}(x) \rangle _g $ is independent of the
position $x$, and hence the summation over the position
$x$ can be trivially done. Moreover all the puncture operators
$P^{ijk}$ are not linearly independent
in the correlation functions, and they are equal to
certain linear combinations of the physical puncture operators
$P^i\ (i=1,\cdots,n)
\ (n\leq M)$\refmark{\FHK}. Thus the equation \prt\ is expressed as
$$
\eqalign{
Z_g^{(1)}&=N  C_l \langle P^l \rangle _g,\cr
\delta C_l&\equiv \delta C_{ijk} A^{ijk}_l,
}
\eqn\parone
$$
where $N$ is the total number of the triangles on the triangulated
surface,
and $A^{ijk}_l$ are the projections to the physical puncture
operators.
We did not change the metric tensor $g^{ij}$. This is because the
change
in the metric tensor will change only the definition of $\delta C_i$
in \parone\ and it will not affect the discussions below.

It is not convenient
to take the thermo-dynamical limit $N\rightarrow\infty$ of
the perturbed partition function $Z^T_g+Z^{(1)}_g$,
as can be seen from the higher orders.
The $p$-th order is a summation over all the selections of $p$
triangles
among the $N$ triangles on the surface. The leading term in $N$
comes
from the cases where any of the $p$ triangles are not next
neighbors. Since
the correlation functions are independent of their positions if they
are not next neighbors, we obtain
$$
Z_g^{(p)}={N^p \over p!} \langle (\delta C_lP^l)^p \rangle
                    +O(N^{p-1}(\delta C)^p).
$$
Thus the partition function is obtained in a compact way as
$$
\eqalign{
Z_g&=Z^T_g+\sum_{p=1}^\infty Z_g^{(p)} \cr
&=\langle \exp (N \delta C_i P^i) \rangle _g
+ ({\rm lower\ orders }). \cr
}
\eqn\cluspart
$$
One can show iteratively that the partition function can be generally
expressed in the form
 $\langle \exp (W) \rangle_g$ in a perturbative treatment,
and that $N \delta C_i P^i$ in \cluspart\
is the first order of a cluster expansion of $W$.
The $p$-th order of the cluster expansion comes from $p$ neighboring
triangles. Thus the terms higher than the first one depend
generally on the lattice structure, and hence they are not universal.
In this paper we assume that $W$ can be treated perturbatively in
$\delta C_{ijk}$.

The calculation rules in the two-dimensional
lattice topological field theories are
very simple in the standard basis\refmark{\FHK}.
The physical puncture operators satisfy the following rules :
$$
\eqalign{
P^iP^j&\sim \delta^{ij} P^i\ \ ({\rm Fusion\ Rule}),\cr
v^i&\equiv \langle P^i \rangle_0, \cr
H&=\sum_{i=1}^n {P^i \over v^i}, \cr
\langle {\cal O} \rangle_g&=\langle {\cal O} H^g \rangle_0, \cr
}
$$
where $\langle {\cal O} \rangle_g$ denotes the correlation function
of operator ${\cal O}$ on a surface of genus $g$.
And the $v^i$'s take only discrete vales, i.e., squares of
integers\refmark{\BP}.
Using the above rules, the numerical value of the partition
function \cluspart\ is
$$
\eqalign{
Z_g&=\sum_{p=0}^\infty Z_g^{(p)}\cr
&=\sum_{i=1}^n \langle P^i \rangle_g \exp (N \delta C_i), \cr
\langle P^i \rangle _g &= (v^i)^{1-g}. \cr}
\eqn\partexpd
$$
Thus in the thermo-dynamical limit $N\rightarrow\infty$ the free
energy per triangle is
$$
\eqalign{
f&=-\lim_{N\rightarrow\infty} {1\over N} \log (Z_g) \cr
&=-\max_{i}\delta C_i}.
$$
This result shows that the position of the lattice topological
field theory
$(\delta C_i=0)$ is a multi-critical first order phase transition
point
with multiplicity $n(n-1)/2$, around which there are $n$ different
phases.

The order parameters of the phase transitions are given by the
one-point
functions of the puncture operators. Define the one-point functions
as follows:
$$
\langle P^l \rangle^{\delta C}_g\equiv {A^l_{ijk} Z(P^{ijk})_g
\over Z_g},
$$
where $Z(P^{ijk})_g$ is the partition function of the triangulated
surface
with a boundary of a triangle with indices $i,j,k$ on its edges.
In the same way as the above calculations for the partition
function, one obtains in the first order of the cluster expansion
$$
\langle P^l \rangle^{\delta C}_g = {\sum_{i=1}^n (\delta^{li}
+O(\delta C)^{li})\langle P^i \rangle_g \exp
(N\delta C_i )
\over \sum_{i=1}^n \langle P^i \rangle_g \exp (N \delta C_i)}.
$$
Here the $O(\delta C)^{li}$ terms are the first order terms of the
cluster expansion for the puncture operator, which
depend on the local lattice structure.
If we are in the $m$-th phase, i.e., $\delta C_m$ is larger than
the others,
we obtain in the thermo-dynamical limit
$$
\langle P^l \rangle^{\delta C}_g = \delta ^{ml}
+O(\delta C)^{ml},
$$
which shows they are good order parameters.

As can be seen in the above calculations, the one-point functions of
the puncture operators of the lattice topological field theory
do not appear anywhere after taking the thermo-dynamical
limit. But they appear as the finite size effect, and in the $m$-th
phase,
the asymptotic form of the free energy per triangle is obtained from
\partexpd\ as
$$
f\sim -\delta C_m -{1 \over N} \log \langle P^m \rangle _g.
$$
One would notice that the coefficient and the topology dependence
of the finite size effect is calculable by the topological field
theory, and that the $1/N$ finite size effect appears only when
$g\neq1$.

Next we will discuss the renormalization group flow near the
multi-critical point.
A partition function must be invariant under a renormalization.
Suppose that the number of triangles $N$ change to $N'$ under
one renormalization step.
The most natural way to realize the invariance of the partition
function
\cluspart\ under the renormalization step is to define the
renormalized
coupling $\delta C'_i$ such that  $N\delta C_i=N'\delta C'_i$.
This renormalization group flow agrees with that of
the Nienhuis-Nauenberg criterion\refmark{\NN},
which gives the renormalization group flow near a discontinuous
fixed point,
and shows that the multi-critical point is the ultraviolet fixed
point of the flow.

Simple examples of such topological fixed points\refmark{\BP} can
be found in the parameter space of the ferromagnetic Potts model.
Consider Potts spins $i,j,\cdots \in \{ 1,\cdots,Q\}$ located on
the sites
of a triangular lattice and interacting among nearest neighbors.
This model can be defined by a metric and a rank-three tensor as
follows:
$$
\eqalign{
g^{(ij)(kl)}&=\delta^{il}\delta^{jk} W(i,j), \cr
C_{(ij)(kl)(mn)}&=\kappa
\delta_{jk}\delta_{lm}\delta_{ni} W(i,j)W(k,l)W(m,n), \cr
W(i,j)&=\exp \left( {\delta_{ij}-1 \over 3T}\right), \cr
}
$$
where $\kappa$ is a constant and $T$ is the temperature.
Those tensors satisfy the topological condition \con\ at $T=0$ and
$T=\infty$
by choosing $\kappa$ appropriately.
At $T=0$ the model is a lattice topological field theory of $n=Q$
and $v^i=1\ (i=1,\cdots,Q)$.
At $T=\infty$ the model is of $n=1$ and $v^1=Q^2$, and the partition
function depends on the topology of the surface.
At the both points the correlation length is zero, and the lattice
topological field theories are realized on the
zero-correlation-length fixed points.

In this letter we clarified some universal natures of the topological
fixed points by the perturbative treatment.
A fixed point on a first order phase transition surface is a
zero-correlation-length fixed point. In fact we have shown that
a lattice topological field theory with more than one physical
observable is on a multi-critical first order phase transition
point.
Any point near a first order phase transition surface would be
transformed
to the neighborhood of a fixed point by a renormalization group
transformation.
{}From this view point we believe that this letter suggests
the possibility that lattice topological field theories would be
useful to understand and classify the first order phase transitions.
Thus it would be an interesting challenge to investigate whether any
lattice topological field theories can also describe the non-trivial
first order phase transition points of the Potts models with $Q>4$.

\ack
We would like to thank H.~Kawai, T.~Yukawa, Y.~Kitazawa,
K.I.~Izawa and N.~Kawamoto for valuable
discussions, comments, criticism and encouragement.
We would like to thank also the members of the elementary particle
sections of
Tokyo University and Kyoto University for their hospitality.
And we would like to thank H.~Kawai for reading this manuscript as
well as N.~Kawamoto for checking the manuscript.

\refout

\bye